# SAR Imaging of Moving Targets via Compressive Sensing

**Jun Wang, Gang Li, Hao Zhang, Xiqin Wang**

Department of Electronic Engineering, Tsinghua University, Beijing 100084, China

Emails: jun-wang05@mails.tsinghua.edu.cn, {gangli, haozhang, wangxq_ee}@tsinghua.edu.cn

## Abstract

An algorithm based on compressive sensing (CS) is proposed for synthetic aperture radar (SAR) imaging of moving targets. The received SAR echo is decomposed into the sum of basis sub-signals, which are generated by discretizing the target spatial domain and velocity domain and synthesizing the SAR received data for every discretized spatial position and velocity candidate. In this way, the SAR imaging problem is converted into sub-signal selection problem. In the case that moving targets are sparsely distributed in the observed scene, their reflectivities, positions and velocities can be obtained by using the CS technique. It is shown that, compared with traditional algorithms, the target image obtained by the proposed algorithm has higher resolution and lower side-lobe while the required number of measurements can be an order of magnitude less than that by sampling at Nyquist sampling rate. Moreover, multiple targets with different speeds can be imaged simultaneously, so the proposed algorithm has higher efficiency.

## 1. Introduction

Synthetic aperture radar (SAR) has been widely used for stationary scene imaging via matched filtering and discrete Fourier transform (DFT). Recently, moving target imaging has attracted much attention. However, the image of moving targets in the observed scene may be displaced and blurred in the azimuth direction because of the motion of targets. Many algorithms have been proposed to deal with this problem. One class of algorithms is based on the estimation of target motion parameters [1-9], e.g., the Doppler rate and the Doppler frequency centroid which are related to the velocities along the range and azimuth directions, respectively. Another class of algorithms combines data reformatting with high order Doppler history analysis, i.e., the Keystone transform [10], the second Keystone transform [11] or the Doppler Keystone transform [12] is used to correct the range cell migration of target, and then the time-frequency analysis [20] or polynomial phase analysis [21-22] is used to compensate the high order terms in the Doppler history. All the algorithms above convert moving target imaging problem to static target imaging problem by compensating the error caused by the target motion, and then the algorithms for stationary scene imaging, e.g., the Range- Doppler algorithm, can be used for imagery formation. Since the Range-Doppler algorithm is based on matched filtering and DFT, the resolutions of final image in the range and azimuth directions are limited by the bandwidth of transmitted signal and the length of the synthetic aperture, respectively. And high side-lobes arise due to the DFT window effect. Moreover, the required Nyquist-rate sampling may cause huge data amount to achieve high resolution. In addition, in above algorithms it is needed to deal with different targets respectively when there are multiple targets with different speeds.

Recently, the compressive sensing (CS) theory [13-15] has been used in a variety of areas for its excellent performance on reconstruction of sparse signals. Consider an equation $\mathbf{y} = \mathbf{\Phi x}$, where $\mathbf{x}$ is an unknown $k$-sparse signal of length $N$ ($k$-sparse means that $\mathbf{x}$ has $k$ large elements), $\mathbf{\Phi}$ is a $M \times N$ measurement matrix, and $\mathbf{y}$ is the measurement vector of length $M$. The CS theory states that $\mathbf{x}$ can be reconstructed by $M = O(k \log N)$ measurements with high probability [13-15]. CS has already been applied to SAR imaging of static targets [16-18]. However, the algorithms in [16-18] are not suitable for SAR imaging of moving targets due to the motion-induced phase error.

In this paper, we propose an algorithm of SAR imaging of moving targets based on the CS technique. We use the matching pursuit strategy to formulate the SAR imaging problem as a basis sub-signal selection problem, where the basis sub-signals are produced by discretizing the extended target space (here the extended target space is defined as a four-dimensional domain spanned by range and azimuth positions and range and azimuth velocities) and synthesizing the SAR data for every discretized spatial position and velocity candidate in the extended target space. There are two assumptions we make here. First, all targets maintain uniform motion in the observation time. Second, the discretized target space is sparse, i.e., the number of targets is much less than the size of discretized target space. By using the CS technique, the spatial positions and the velocities of the targets can be reconstructed with a relatively small number of random measurements. Compared with traditional DFT-based algorithms of moving target imaging, the proposed algorithm can obtain higher resolution and lower side-lobe with fewer measurements. Moreover, multiple targets with different speeds can be simultaneously imaged by using our algorithm and thus the imaging efficiency can be improved.

The remainder of this letter is organized as follows. The SAR signal model of moving targets is discussed in Section 2. The proposed algorithm is formulated in Section 3. Numerical examples are provided in Section 4.

*Notation*: Vectors and matrices are denoted by boldface letters; all vectors are column vectors; $(\cdot)^T$ denotes the transpose operation; $\|\cdot\|_1$ and $\|\cdot\|_2$ denotes $l_1$ and $l_2$ norms, respectively; $(\cdot)^{vec}$ generates a column vector by stacking the columns of a matrix one underneath the other in sequence, e.g., for a $M \times N$ matrix $\mathbf{X}$,

$$\mathbf{X}^{vec} \triangleq [\underbrace{\mathbf{X}(1,1), \mathbf{X}(2,1), \cdots, \mathbf{X}(M,1)}_{\text{1st column of } \mathbf{X}}, \underbrace{\mathbf{X}(1,2), \mathbf{X}(2,2), \cdots, \mathbf{X}(M,2)}_{\text{2nd column of } \mathbf{X}}, \cdots, \underbrace{\mathbf{X}(1,N), \mathbf{X}(2,N), \cdots, \mathbf{X}(M,N)}_{N\text{th column of } \mathbf{X}}]^T$$

## 2. SAR signal model of moving targets

The typical side-looking SAR geometry is illustrated in Fig. 1. *X*-axis and *Y*-axis are range and azimuth directions, respectively. The radar platform flies along *Y*-axis with constant speed $v$, and a point target T moves with constant range speed $v_x$ and azimuth speed $v_y$, which lies in $(x_t, y_t)$ at $\eta = 0$, here $\eta$ is the azimuth time (or slow time). Suppose that the radar platform lies in $(0,0)$ at $\eta = 0$.

Assume that the radar transmits a chirp signal, then the baseband echo from T is:

$$b_t(\tau,\eta) = \sigma_t \cdot s_t(\tau,\eta) = \sigma_t \cdot \omega_r(\tau - 2R(\eta)/c) \cdot \omega_a(\eta - \eta_c) \cdot \exp[j\pi K_r(\tau - 2R(\eta)/c)^2] \cdot \exp[-j4\pi f_0 R(\eta)/c], \quad (1)$$

where $\tau$ is the range time (or fast time), $\omega_r$ is the range envelope which is determined by the transmitted signal, and $\omega_a$ is the azimuth envelope which is determined by the beam of the radar antenna, $f_0$ is the carrier frequency, $\sigma_t$ is the reflectivity of T, $s_t(\tau,\eta)$ corresponds to the received echo from a target which has the same position and speed with T while the reflectivity is 1. $\eta_c$ is the zero-Doppler time of the target which can be expressed as $\eta_c = \dfrac{y_t}{v - v_y}$, $R(\eta)$ is the instantaneous distance from the radar to the target and it can be expressed as

$$R(\eta) = \sqrt{(x_t + v_x\eta)^2 + (y_t + (v_y - v)\eta)^2}, \quad (2)$$

An approximate expression of (2) can be obtained by using the first two terms in the Taylor expansion at $\eta = \eta_c$:

$$R(\eta) \approx x_t + v_x(\eta - \eta_c) + \frac{(v_y - v)^2}{2x_t}(\eta - \eta_c)^2, \quad (3)$$

We perform range compression on (1) and then the signal can be expressed as

$$b_{rc}(\tau,\eta) \approx \sigma_t \cdot p_r(\tau - 2R(\eta)/c) \cdot \omega_a(\eta - \eta_c) \cdot \exp\{-j\frac{4\pi f_0}{c}[x_t + v_x(\eta - \eta_c) + \frac{(v_y - v)^2}{2x_t}(\eta - \eta_c)^2]\}, \quad (4)$$

where $p_r(\cdot)$ denotes the profile of the range-compressed signal, for example, if $\omega_r$ is a rectangular envelope, then $p_r$ is a sinc function.

It is clear that, 1) the range cell migration may occur if the variance of $R(\eta)$ during the observation duration is larger than the range resolution; 2) the last term of the phase in (4) induces the time-varying Doppler frequency and will cause the final image blurred. The goal of motion compensation is to correct both of the range cell migration and the time-varying Doppler frequency. Once this is done, the focused image of the target can be directly obtained by performing the Fourier transform in terms of $\eta$.

From (1) we can express the SAR echo from the observed scene as

$$b(\tau,\eta) = \iint \sigma(x,y) \cdot s_{x,y}(\tau,\eta) dx dy , \qquad (5)$$

where $s_{x,y}(\tau,\eta)$ is the received echo from the target with reflectivity 1 located at $(x,y)$ when $\eta = 0$ (see(1)), $\sigma(\cdot)$ denotes the reflectivities of possible targets. That is to say, if there is a target located at $(x,y)$ when $\eta = 0$, $\sigma(x,y)$ denotes the reflectivity of this target; if there is no target at the position $(x,y)$ when $\eta = 0$, $\sigma(x,y) = 0$. Here, we define $\sigma(\cdot)$ as reflectivity profile. In (5) $(x,y)$ constitutes the target space $\pi_T$ which lies in the product space $[x_i, x_f] \times [y_i, y_f]$. Here $(x_i, y_i)$ and $(x_f, y_f)$ denotes the initial and final positions of the target space of interest along each axis, respectively. The goal is to retrieve the profile $\sigma(\cdot)$ that indicate the reflectivities and the positions of all possible targets (i.e., the image of the target space) using the measurements $b(\tau,\eta)$.

Normally we work with the discrete version of (5). If the measurements and the image (i.e., the reflectivity profile $\sigma(\cdot)$) can be expressed as vectors, a discretized reflectivity profile $\boldsymbol{\sigma}$ is related to the sampled measurement vector $\mathbf{b}$ through a matrix. However, $s_{x,y}(\tau,\eta)$ can not be determined because the speed of the target located at $(x,y)$ is unknown. A method to solve this problem is to increase the dimensions of the target space. Consider the extended target space $\pi_{T,e}$ which lies in the product space $[x_i, x_f] \times [y_i, y_f] \times [v_{x,i}, v_{x,f}] \times [v_{y,i}, v_{y,f}]$. Here $(v_{x,i}, v_{y,i})$ and $(v_{x,f}, v_{y,f})$ denotes the minimum and maximum velocity candidates of the extended target space along range and azimuth velocity axis, respectively. Thus the received echo can be expressed as

$$b(\tau,\eta) = \iint \iint \sigma(x,y,v_x,v_y) \cdot s_{x,y,v_x,v_y}(\tau,\eta) dx dy dv_x dv_y , \qquad (6)$$

where $\sigma(\cdot)$ is the reflectivity profile of the extended target space, $s_{x,y,v_x,v_y}(\tau,\eta)$ is the received echo from one target which is located at $(x,y)$ at $\eta = 0$ with reflectivity 1 and whose range and azimuth speeds are $v_x$ and $v_y$, respectively (see(1) and (3)). From (1), (3) and (6) one can see that targets with different positions or speeds provide different contributions on the measurements $b(\tau,\eta)$, which implies that the moving target can be uniquely determined in the extended target space [5]. Once the energy distribution of the extended target space is obtained, the sub-space $(x,y)$ and the sub-space $(v_x,v_y)$ show the positions and the velocities of all possible targets, respectively. In what follows, we describe how to recover the extended target space $\sigma(x,y,v_x,v_y)$ from the measurements $b(\tau,\eta)$.

## 3. CS for SAR moving target imaging

First we discretize the extended target space $\pi_{T,e}$ as a four dimensional space of size $N_1 \times N_2 \times P \times Q$, where $N_1, N_2, P$ and $Q$ are the numbers of discretized values of $x, y, v_x$ and $v_y$, respectively. The bin sizes of $x, y, v_x$ and $v_y$ are $\rho_r, \rho_a, \mu_r$ and $\mu_a$, respectively.

If motion parameters of a moving target match the coordinates $(n_1, n_2, p, q)$ in the discretized extended target space, from (2) we can directly obtain the distance from the radar to the target:

$$R_{(n_1,n_2,p,q)} = \sqrt{(x(n_1)+v_x(p)\cdot\eta)^2 + (y(n_2)+(v_y(q)-v)\cdot\eta)^2}, \tag{7}$$

where $x(n_1) = x_i + \rho_r n_1$, $y(n_2) = y_i + \rho_a n_2$, $v_x(p) = v_{x,i} + \mu_r p$, $v_y(q) = v_{y,i} + \mu_a q$.

Suppose that the reflectivity of the target is 1, and then the baseband echo can be expressed as (see (1) and (6)):

$$\begin{aligned} d_{(n_1,n_2,p,q)}(\tau,\eta) &\triangleq s_{x(n_1),y(n_2),v_x(p),v_y(q)}(\tau,\eta) \\ &= \omega_r(\tau - 2R_{(n_1,n_2,p,q)}(\eta)/c)\cdot\omega_a(\eta-\eta_c)\cdot\exp[j\pi K_r(\tau-2R_{(n_1,n_2,p,q)}(\eta)/c)^2]\cdot\exp[-j4\pi f_0 R_{(n_1,n_2,p,q)}(\eta)/c], \end{aligned} \tag{8}$$

We define (8) as the sub-signal corresponding to the coordinates $(n_1, n_2, p, q)$. Thus, the SAR received echo of the observed scene can be expressed as the sum of all sub-signals (see(6)):

$$b(\tau,\eta) = \sum_{n_1=0}^{N_1-1}\sum_{n_2=0}^{N_2-1}\sum_{p=0}^{P-1}\sum_{q=0}^{Q-1} a(n_1,n_2,p,q)\cdot d_{(n_1,n_2,p,q)}(\tau,\eta), \tag{9}$$

where $a(n_1,n_2,p,q) = \sigma(x(n_1),y(n_2),v_x(p),v_y(q))$. Suppose that $a(n_1,n_2,p,q)$ is the $(n_1,n_2,p,q)$th element of **a**, and thus **a** is a $N_1 \times N_2 \times P \times Q$ reflectivity profile matrix.

Assume that the range and azimuth sample rates are $f_s$ and $f_a$ respectively, the range and azimuth sample numbers are $N_r$ and $N_a$ respectively. Thus, the SAR received echo can be discretized as a $N_r \times N_a$ matrix, and its $(m,n)$th element is

$$b(\tau_m,\eta_n) = \sum_{n_1=0}^{N_1-1}\sum_{n_2=0}^{N_2-1}\sum_{p=0}^{P-1}\sum_{q=0}^{Q-1} a(n_1,n_2,p,q)\cdot d_{(n_1,n_2,p,q)}(\tau_m,\eta_n), \tag{10}$$

where $\tau_m = \tau_0 + m/f_s$, $\eta_n = n/f_a$, $\tau_0 = x_i/(2c)$.

Let $\mathbf{b}$ is $N_r \times N_a$ matrix whose $(m,n)$ th element is $b(\tau_m, \eta_n)$, and let $\mathbf{d}_{(n_1,n_2,p,q)}$ is a $N_r \times N_a$ matrix whose $(m,n)$ th element is $d_{(n_1,n_2,p,q)}(\tau_m, \eta_n)$. Vectorize all $\mathbf{d}_{(n_1,n_2,p,q)}$ for $n_1=0, \ldots, N_1-1$, $n_2=0, \ldots, N_2-1$, $p=0, \ldots, P-1$, $q=0, \ldots, Q-1$ and arrange all of them as a matrix $\mathbf{\Phi}$:

$$\mathbf{\Phi} \triangleq [\mathbf{d}^{vec}_{(0,0,0,0)}, \cdots, \mathbf{d}^{vec}_{(n_1,n_2,p,q)}, \cdots, \mathbf{d}^{vec}_{(N_1-1,N_2-1,P-1,Q-1)}], \qquad (11)$$

where $\mathbf{d}^{vec}_{(n_1,n_2,p,q)}$ is a vector of length $N_r N_a$, which is the vectorization of $\mathbf{d}_{(n_1,n_2,p,q)}$, $\mathbf{\Phi}$ is a $N_r N_a \times N_1 N_2 PQ$ matrix. Thus (9) can be expressed as

$$\mathbf{b}^{vec} = \mathbf{\Phi} \cdot \mathbf{a}^{vec}, \qquad (12)$$

where $\mathbf{a}^{vec}$ and $\mathbf{b}^{vec}$ are vectorizations of $\mathbf{a}$ and $\mathbf{b}$, whose sizes are $N_1 N_2 PQ \times 1$ and $N_r N_a \times 1$, respectively. All $\mathbf{d}^{vec}_{(n_1,n_2,p,q)}$ for $n_1=0, \ldots, N_1-1$, $n_2=0, \ldots, N_2-1$, $p=0, \ldots, P-1$, $q=0, \ldots, Q-1$, are called sub-signals, and thus the solution of (12) can be considered as the sub-signal selection problem, which can be solved by the CS technique.

According to the CS theory [13-15], if $\mathbf{a}^{vec}$ is a k-sparse vector, $M = O[k\log(N_1 N_2 PQ)]$ measurements are enough to reconstruct $\mathbf{a}^{vec}$. Consider $\mathbf{b}^{vec}|_M$ as randomly selected $M$ elements of $\mathbf{b}^{vec}$, i.e., the $m$th element of $\mathbf{b}^{vec}|_M$ is the $\gamma_m$ th element of $\mathbf{b}^{vec}$, where $\gamma_m$ are integers for $m=1,2,\cdots,M$ and $1 \le \gamma_1 < \gamma_2 < \cdots < \gamma_M \le N_r N_a$. $\mathbf{\Phi}|_M$ means a $M \times N_1 N_2 PQ$ matrix whose $m$th row is the $\gamma_m$ th row of $\mathbf{\Phi}$. Then by randomly selecting $M$ elements from $\mathbf{b}^{vec}$ and corresponding $M$ rows from $\mathbf{\Phi}$, (12) can be rewritten as

$$\mathbf{b}^{vec}|_M = \mathbf{\Phi}|_M \times \mathbf{a}^{vec}, \qquad (13)$$

which can be solved by

$$\text{minimize } \|\mathbf{a}^{vec}\|_1, \text{ subject to } \|\mathbf{b}^{vec}|_M - \mathbf{\Phi}|_M \times \mathbf{a}^{vec}\|_2 < \varepsilon, \qquad (14)$$

where $\varepsilon$ is the error threshold. The solution (14) can be obtained by using the CoSaMP algorithm [19]. Then the reflectivity profile $\mathbf{a}$ can be obtained. Thus the correct sub-signals that really contribute to the received SAR echo are selected and the projection coefficients of the SAR echo on these sub-signals are solved. The positions and reflectivities of the targets directly represent the SAR final image, and meanwhile the velocities of the targets are also known. It is clear that $k \ll N_1 N_2$ since the observed scene is sparse, and in general $N_1 N_2 < N_r N_a$, thus $k \ll N_r N_a$. $M = O[k\log(N_1 N_2 PQ)]$, so $M \ll N_r N_a$ which means the measurements in the proposed algorithm are much fewer than that in the traditional algorithm.

## 4. Numerical Examples

In this section, we use simulated data to verify the feasibility and effectiveness of the proposed algorithm. Parameters of the SAR system are shown in Table I. The size of the observed scene is 15m×15m, and the velocity in each direction is limited in (-10m/s,10m/s). The bins of the dimensions of the extended target space are 0.5m, 0.5m, 2m/s and 2m/s, respectively, i.e., $\rho_r = \rho_a = 0.5$m, $\mu_r = \mu_a = 2$m/s, and thus the extended target space is a $31 \times 31 \times 11 \times 11$ four-dimensional space, i.e., $N_1 = N_2 = 31$, $P = Q = 11$. The radar platform transmits chirp signal and suppose that the range and azimuth envelopes are both rectangular envelopes.

### A. Feasibility

Suppose that there are three targets with reflectivity 1 in the observed scene. The first target is static, the second one is moving along range direction with speed 10m/s, and the third one is moving with range speed 4m/s and azimuth speed 4m/s. When $\eta = 0$, the three target are located in (4m, 2.5m), (7.5m, 10m) and (11.5m, 8m) as illustrated in Fig. 2(a).

Firstly, the imaging result by the traditional algorithm is shown. The range sample number $N_r = 1213$ and the azimuth sample number $N_a = 595$, so the measurement number of the traditional algorithm is $N_r N_a$=721735. The motion parameters of the three targets are estimated by the range cell migration correction and the high order Doppler phase compensation as done in [12]. Then the imaging result of the Range-Doppler algorithm is illustrated in Fig. 2(b). We can see that the image is blurred due to the poor resolution and the high side-lobe.

Then the imaging result by the proposed algorithm is shown. Randomly select $M$=100 measurements from the SAR echo $\mathbf{b}^{vec}$, and then solve (14) using the CoSaMP algorithm [19]. The imaging result is illustrated in Fig. 2(c). Compared with Fig. 2(b), the resolution is significantly improved and the side-lobe is removed, meanwhile the measurements are reduced from 721735 to 100. Another point to note is that the traditional algorithm [12] needs to estimate motion parameters of each target one by one for above three targets. Thus, the proposed algorithm is more efficient in the case of multiple targets with different velocities.

### B. Effect of the number of measurements

An important measure of CS technique is the successful recovery probability, plotted in Fig. 3 as a function of the number of measurements for a varying number of targets (sparsity level $P$), i.e., 1-4. The positions and speeds of targets in the observed scene are randomly generated. For each point on the plot, 200 trials of the proposed algorithm are run. Suppose that $\hat{\mathbf{a}}^{vec}$ is the solution of the reflectivity profile $\mathbf{a}^{vec}$ by (14), then if $\left\|\hat{\mathbf{a}}^{vec}-\mathbf{a}^{vec}\right\|_{2}/\left\|\mathbf{a}^{vec}\right\|_{2}<0.1$, the image is counted as a successful recovery. The number of successful recoveries over 200 trials yields the probability of successful recovery (PSR) in terms of the measurement and the sparsity level *P*. It can be observed that increasing the sparsity level *P*, i.e., the number of targets, requires more measurements for the same level of PSR. Moreover, even for *P*=4, 60 measurements are enough to achieve high PSR, which is still far less than the 721735 measurements needed for traditional algorithms.

### *C. Effect of the noise*

To analyze the effect of versus (additive) noise level, the algorithm is applied to the received SAR data with SNRs from -15 to 35dB. The number of the targets is 1 whose position and speed are generated randomly. Fig. 4 shows the PSR as a function of SNR for varying number of measurements. For each point on the plot, 100 trials of the proposed algorithm are run. It can be observed that increasing the number of measurements allows the method to perform at lower SNR values. Even with small number of measurements, i.e., 20, high successful recovery rates can be obtained with SNRs more than 15dB.

## 5. Conclusion

This paper proposes a CS-based algorithm for SAR imaging of moving targets. The received SAR echo is decomposed as the sum of many basis sub-signals that are generated by discretizing the target spatial domain and velocity domain and synthesizing the SAR received data for every discretized spatial position and velocity candidate. By using the CS technique, the correct sub-signals that contribute to the received SAR echo are selected and the projection coefficients of the SAR echo on these sub-signals are solved. The positions and reflectivities of the targets directly represent the SAR final image, and meanwhile the velocities of the targets are also obtained. Numerical examples show that, compared with traditional algorithms based on Doppler phase analysis and DFT, the proposed algorithm can obtain better imagery quality with higher resolution and lower side-lobe. Moreover, the measurements used by the proposed algorithm are much fewer. In addition, multiple targets with different velocities can be imaged simultaneously and thus the proposed algorithm has higher efficiency**.**

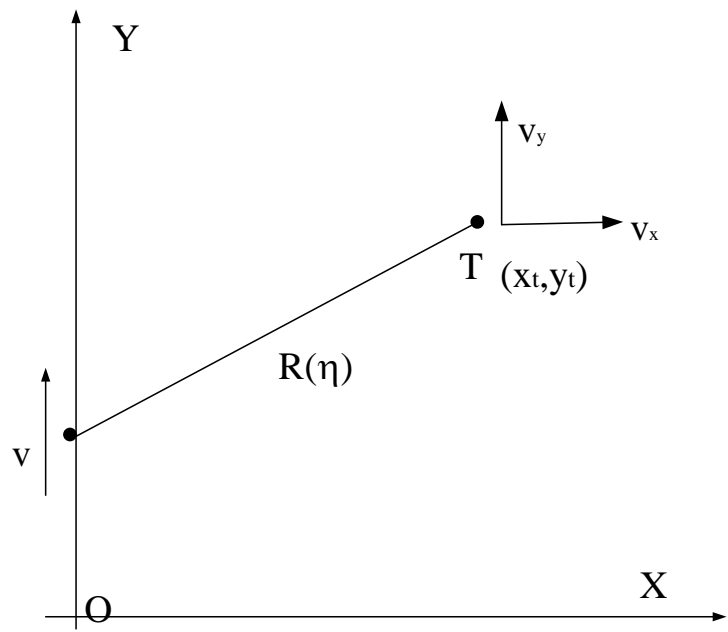

Fig. 1. Slant plane of SAR geometry.

11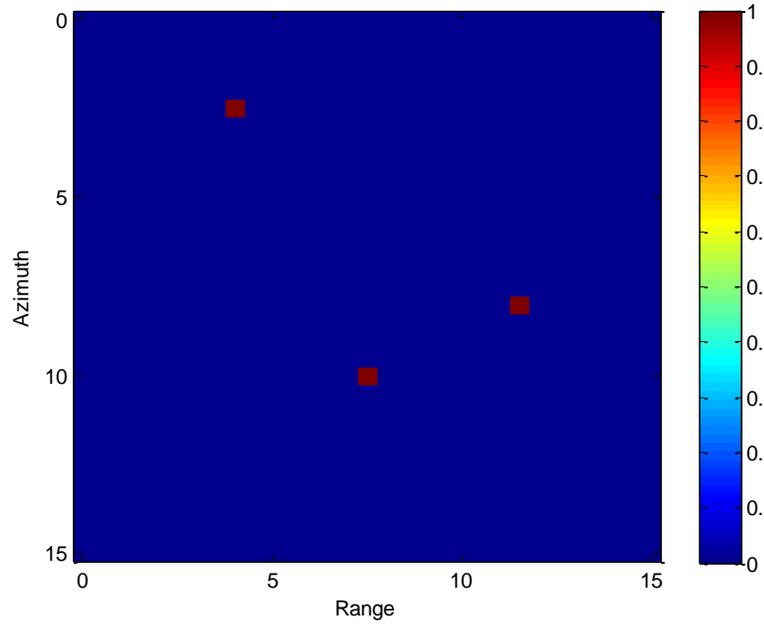

(a)

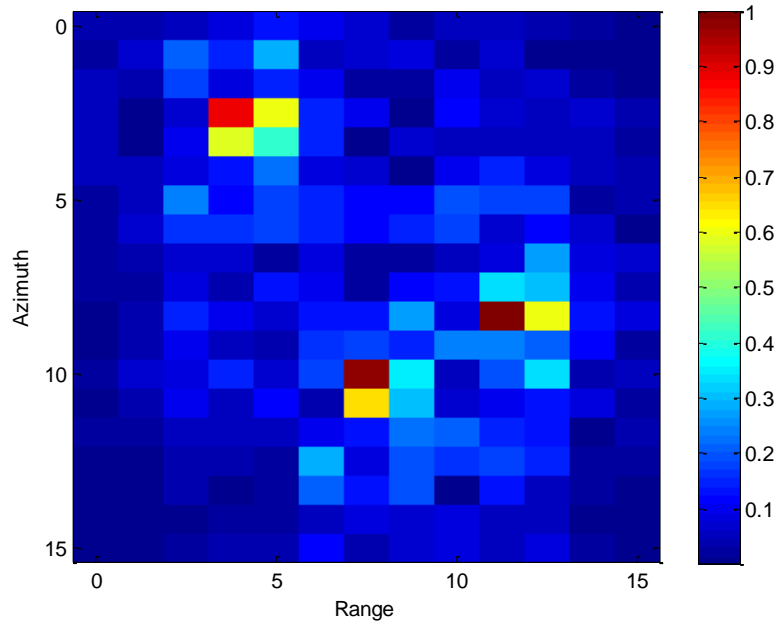

(b)

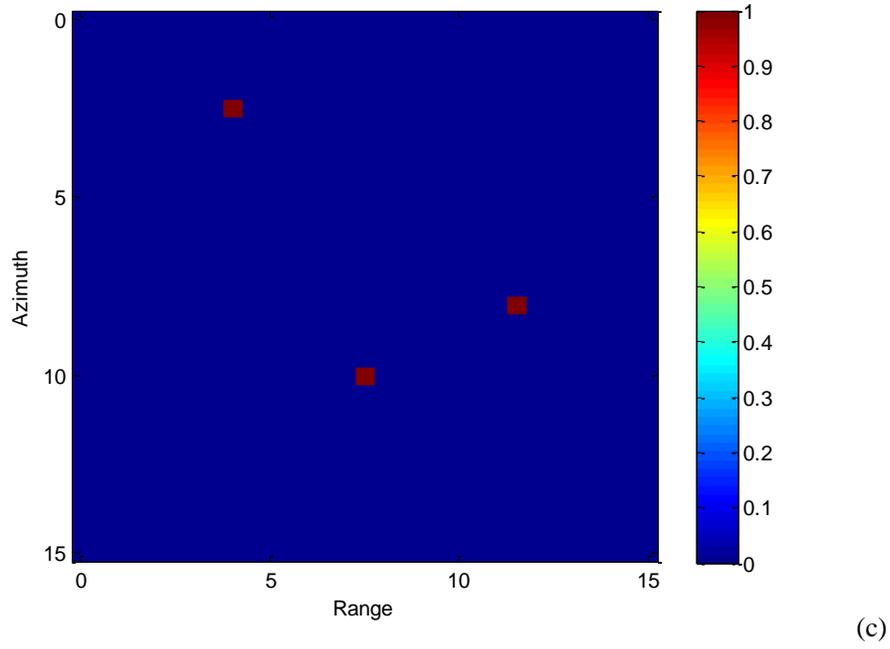

(c)

Fig. 2. (a)The observed scene; (b) The imaging result by the traditional DFT-based algorithm; (c) The imaging result by the proposed algorithm.

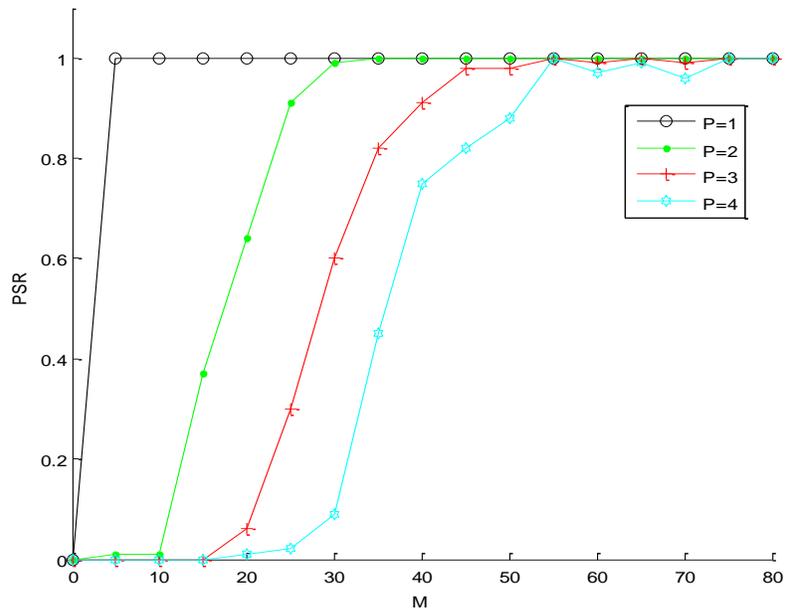

Fig. 3 The probability of successful recovery (PSR) vs. the number of measurements ($M$) for varying number of targets ($P$) in the target space.

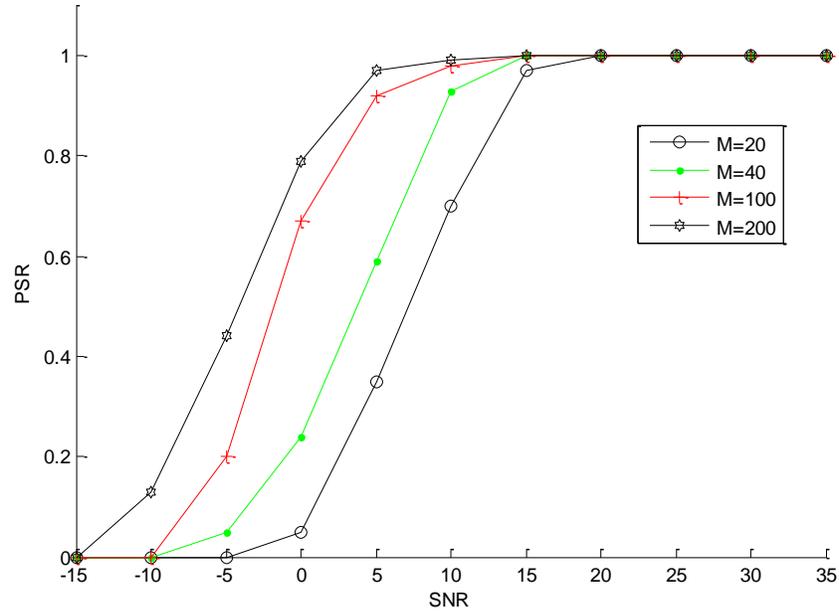

Fig. 4 The probability of successful recovery (PSR) vs. SNR for varying number of measurements *M*

TABLE I

PARAMETERS OF THE SAR SYSTEM

| Scene center range | 30km | Platform speed | 250m/s |
|---|---|---|---|
| Pluse width | 10us | Carrier frequency | 9.375GHz |
| Transmitted signal bandwidth | 100MHz | Antenna length | 2m |
| Radar sampling rate | 120MHz | Radar wavelength | 0.032m |
| Pulse repitition frequency | 300Hz | | |